\documentclass[%
 reprint,
 amsmath,amssymb,
 aps,
pre]{revtex4-2}

\usepackage{blkarray}
\usepackage{amsmath}
\usepackage{graphicx}
\usepackage{dcolumn}
\usepackage{bm}
\usepackage{hyperref}
\usepackage{xcolor}

\begin{document}


\title{Temperature-Noise Interplay in a Coupled Model of Opinion Dynamics}

 \author{Anna Chmiel and Julian Sienkiewicz}
 \affiliation{Faculty of Physics, Warsaw University of Technology, Koszykowa 75, 00-662 Warszawa, Poland}
 \email{julian.sienkiewicz@pw.edu.pl}






\begin{abstract}
We consider a coupled system mimicking opinion formation under the influence of a group of $q$ neighbors ($q$-lobby) that consists of an Ising part governed by temperature-like parameter $T$ and a voter dynamics parameterized by noise probability $p$ (independence of choice). Using rigorous analytical calculations backed by extensive Monte Carlo simulations, we examine the interplay between these two quantities. Based on the theory of phase transitions, we derive the relation between $T$ and $p$ at the critical line dividing the ordered and disordered phases, which takes a very simple and generic form $T(p-a)=b$ in the high temperature limit. For specific lobby sizes, we show where the temperature and noise are balanced, and we hint that for large $q$, the temperature-like dynamics prevails.  
\end{abstract}

\maketitle

\section{Introduction}
In spite of a multitude of different opinion dynamics models \cite{Castellano2009,Jusup2021}, the notion of social temperature remains elusive. Although it can be intuitively understood as a measurement that quantifies the extent to which a group's average opinion is susceptible to changing \cite{Bahr1998}, degree of randomness in the behavior of individuals \cite{Kacperski1999} or willingness to change one’s opinion \cite{Korbel2023}, no universal definition thereof can be found.

Additionally, in this context, the (social) temperature is often interchangeably used in an explicit way with the idea of uncertainty \cite{Galesic2021}, attention or lack thereof \cite{Sznajd-Weron2024,Galesic2021}, (social) noise level \cite{Holyst2023} and randomness or pure noise \cite{Castellano2009}. It has also been argued that \textit{independence} plays a role similar to temperature \cite{Nyczka2012,Jedrzejewski2019}

In this paper, we aim to find a bridge between these concepts. To address this task, we use two very similar approaches that examine the influence of a selected group of agents (a q-lobby) on the opinion of an individual. However, one of these approaches, called the q-neighbor Ising model (for brevity further called q-Ising model) \cite{Jedrzejewski2015}, uses the concept of (pseudo)temperature, while the second one -- the q-voter model \cite{Nyczka2012} utilizes the idea of independence of choice that could be understood as an intrinsic noise. In the same way as the original Ising and voter models have found their way into different areas of science \cite{Macy2024,Fernndez-Gracia2014}, their versions imitating lobby influence have also created new paths to follow in statistical physics \cite{Park2017,Aiudi2023} and social sciences that touch such phenomena as polarization \cite{Jacob2023} contrarian effect \cite{Iacomini2023} or conformity bias \cite{Denton2021}.

Apart from exhibiting interesting statistical physics phenomena (e.g., first and second order phase transitions, tricriticality), both models in question are heavily influenced by social psychology research on the impact of conformity. In particular, we refer here to Solomon Asch experiments \cite{Asch1955,Asch1956} that have shown this impact to increase with the size of the unanimous influence group, but only up to a certain threshold; moreover, the conformity is reduced in a drastic manner if the group of influence is not unanimous. The probability $p$ in the q-voter model and the temperature-like parameter $T$ in its q-Ising counterpart can be used to tune the level of independence. However, individuals typically live simultaneously in different communities -- be it offline or online -- that can be treated as different levels in a multilayer network, where they are subject to opinion change suggestions. Such a setting has been investigated in the form of a duplex q-voter \cite{Chmiel2015} and duplex q-Ising models \cite{Chmiel2017} that have proven that the addition of a level dramatically changes the behavior of the system, e.g., while for the monolayer q-Ising we have second or first order phase transitions between the ordered and disorder phase depending on the lobby size, in the q-Ising duplex only second order transition takes place.

In this study, we couple these two different generic dynamics and we examine the conditions that are necessary for one of the models to prevail, thus finding the relation between the temperature and noise (independence). Additionally, owing to the parametrization of the model with the lobby size $q$, we show how the number of active neighbors affects the above relation and alters the system's behavior.

\begin{figure}[!ht]
    \includegraphics[width=\columnwidth]{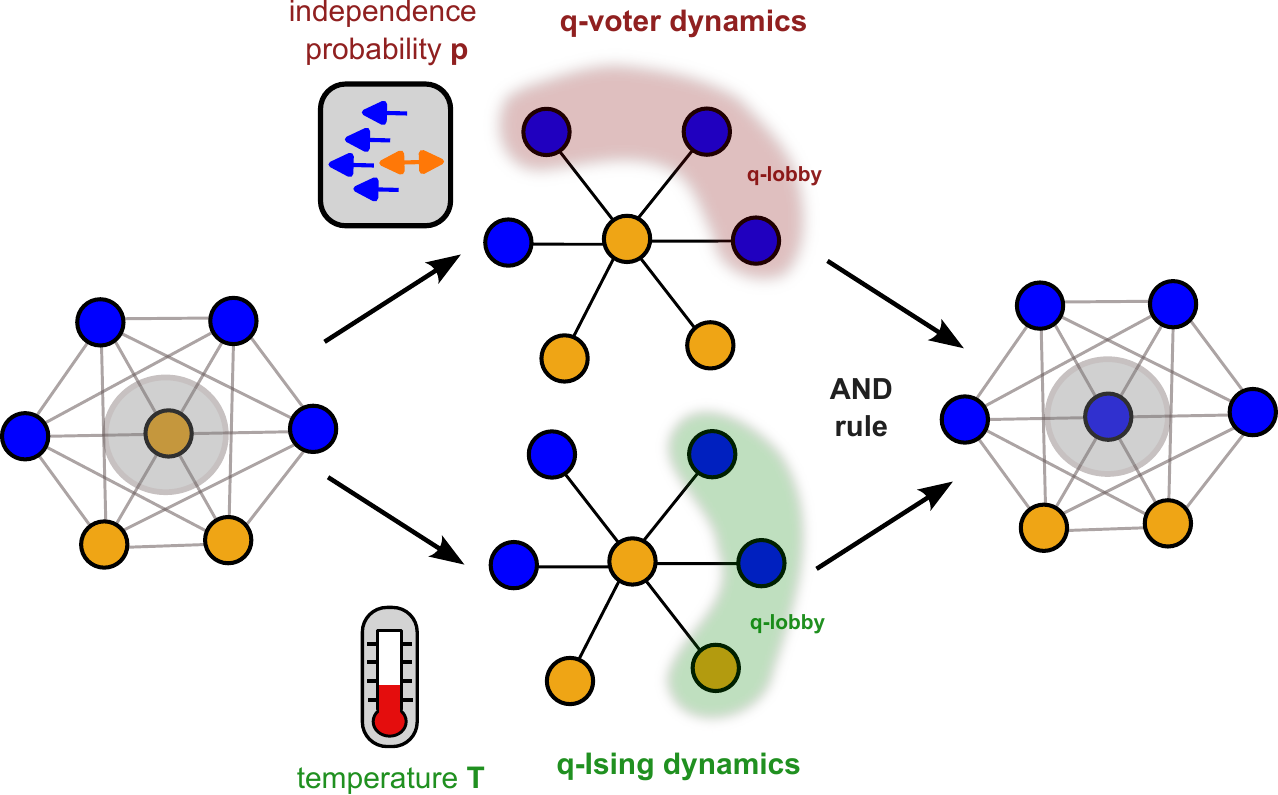}
    \caption{Schematic representation of the examined model}
\label{fig:fig1}
\end{figure}

\section{Model description and its fundamental properties}\label{sec:model}

The system consists of $N$ agents, each holding a binary opinion $s_i=\pm1$, acting on top of a complete graph, described by total magnetization $m=\frac{1}{N}\sum_{i=0}^{i=N}s_i$ (average spin per node). In each time step, one agent is selected at random, and outcomes of two types of dynamical rules performed independently (q-voter and q-Ising, described further) are checked (see Fig.~\ref{fig:fig1}). If both dynamics indicate a change of opinion, the selected node flips its state -- it is a so-called AND rule, previously used in threshold dynamics \cite{Lee2014}, majority-vote models \cite{Choi2019}, and q-voter \cite{Chmiel2015,Gradowski2020} and q-Ising dynamics \cite{Chmiel2017}. Alternatively, one can describe the system as a duplex (two-layer multiplex network) with different dynamics on each level and the condition that the state of node has to be the same in both layers, which in the case of opinion dynamics, can be understood as ``no hypocrisy rule'', i.e., the agent presents the same opinion in each community it belongs to. Different types of coupling may be considered, e.g., the OR rule, where it suffices to prompt the change on one level to flip the node. The AND rule has been chosen owing to the possibility of comparing with previous studies connected to duplex networks \cite{Chmiel2015,Chmiel2017,Chmiel2020}, but also to imitate the fact that acquiring the same opinion from more than one group reinforces it and allows it to last (in a similar way as it is done in the case of echo chamber effect for the spread of information \cite{Cinelli2021,Baumann_2020,Gajewski2022}). In contrast, the OR rule should lead to fast suppression of the ordered phase (see the Discussion section for a detailed derivation of the rate equation for the OR rule and the description of differences between OR and AND rules). 

The key component of both types of dynamics is the \textit{q-lobby} (q-panel) -- a set of $q$ nodes selected at random from all neighbors of the given agent $i$ (here, given the complete graph, out of all nodes). In the case of the q-voter dynamics \cite{Nyczka2012}, the agent flips its state only when the $q$-lobby is homogeneous, i.e., all the chosen neighbors have the same state, however, it still has the option to behave (i.e., change its opinion) independently, which takes place with probability $p$, called later probability of independence. Thus, in the infinite system, the probability $\gamma^{+}_v$ of an increase in the magnetization for the q-voter dynamics reads 
\begin{equation}
\gamma^{+}_v \equiv \gamma^{+}_v(q,m,p)=2^{-q}(1-p)(1+m)^{q}+\frac{p}{2}.    
\end{equation}
In the same manner, owing to the symmetry of the problem $\gamma^{-}_v \equiv \gamma^{+}_v(q,-m,p)$. In the case of the q-Ising model \cite{Jedrzejewski2015}, in analogy to the original Ising model, the opinion of the agent is flipped with probability $\min\{1,\exp(\Delta E/T)\}$, where $T>0$ is temperaturelike parameter, $\Delta E = 2s_i\sum_{nn}s_j$ and summation goes over all nodes in the q-lobby. Thus the probability $\gamma^{+}_I$ of an increase in the magnetization for the q-Ising dynamics reads
\begin{equation}
\gamma^{+}_I\equiv \gamma^{+}_I(q,m,T)=2^{-q}\sum_{k=0}^q\binom{q}{k}(1+m)^{q-k}(1-m)^k E_{qk},
\end{equation}
where $E_{qk}=\min\{1,\exp[2(q-2k)/T]\}$. Similarly, $\gamma^{-}_I\equiv \gamma^{+}_I(q,-m,T)$.

Choosing the topology of a complete graph over a specific network model might seem an oversimplification; however, judging from previous works on q-voter and q-Ising network duplexes \cite{Gradowski2020,Krawiecki2023}, the character of the observed phenomena should stay the same.

Given the above rules and the condition that both dynamics need to suggest the flip, the rate equation $F_q \equiv F(q,m,p,T)$ governing the coupled system reads
\begin{equation}
    F_q = \frac{1-m}{2}\gamma^{+}_I\gamma^{+}_{v}
    -\frac{1+m}{2}\gamma^{-}_{I}\gamma^{-}_{v} = 0.
\label{eq:rate}
\end{equation}

The standard procedure to obtain the critical temperature (or critical value of noise) for systems characterized by continuous phase transition or the lower spinodal for the discontinuous one is to require that 
\begin{equation}
\left(\frac{\partial F_q}{\partial m}\right)_{m=0}=0    
\end{equation}
and use it to obtain the relevant value of $T_c$ or $p_c$. Here, we follow the same procedure; however, it leads to a relation $f(p,T)=0$ instead.  As $F_q$ is linear with respect to $p$ for any value of $q$, a straightforward solution presenting the critical value of the noise as a function of temperature reads
\begin{equation}
p^*_q(T)=\frac{\sum\limits_{k=0}^{q}\tbinom{q}{k}\left(2q-2k-1\right)E_{qk}}{\sum\limits_{k=0}^{q}\tbinom{q}{k}[(2^{q-1}-1)(1+2k-q)+q]E_{qk}}.
\label{eq:sep1}
\end{equation}
which it is equivalent of solving $m(T,p)=0$ when an explicit form of $m(T,p)$ is given.

Owing to the very form of Eq. (\ref{eq:rate}), i.e., a sum of $\exp(a\beta)$ with $\beta=T^{-1}$, linear expansion $p^*_q(\beta)$ around $\beta=0$ yields 
\begin{equation}
    p^*_q(T) \underset{T \gg1}\approx a_q + b_qT^{-1}
\label{eq:pq}
\end{equation}
or, equivalently, $T(p-a_q)=b_q$. In the above equation, $a_q$ equals the critical value of $p$ for a monolayer q-voter model as it is equivalent to the $T \to \infty$ case. The above formula holds true for any value of $q$ considered in this study and stands as a first key point we deliver in this paper, showing a concise relation between the temperature and the noise in the high temperature limit at the critical line with $b_q$ expressing the way $T^{-1}$ relates to noise $p$. 

\begin{figure*}[!ht]
         \includegraphics[width=\textwidth]{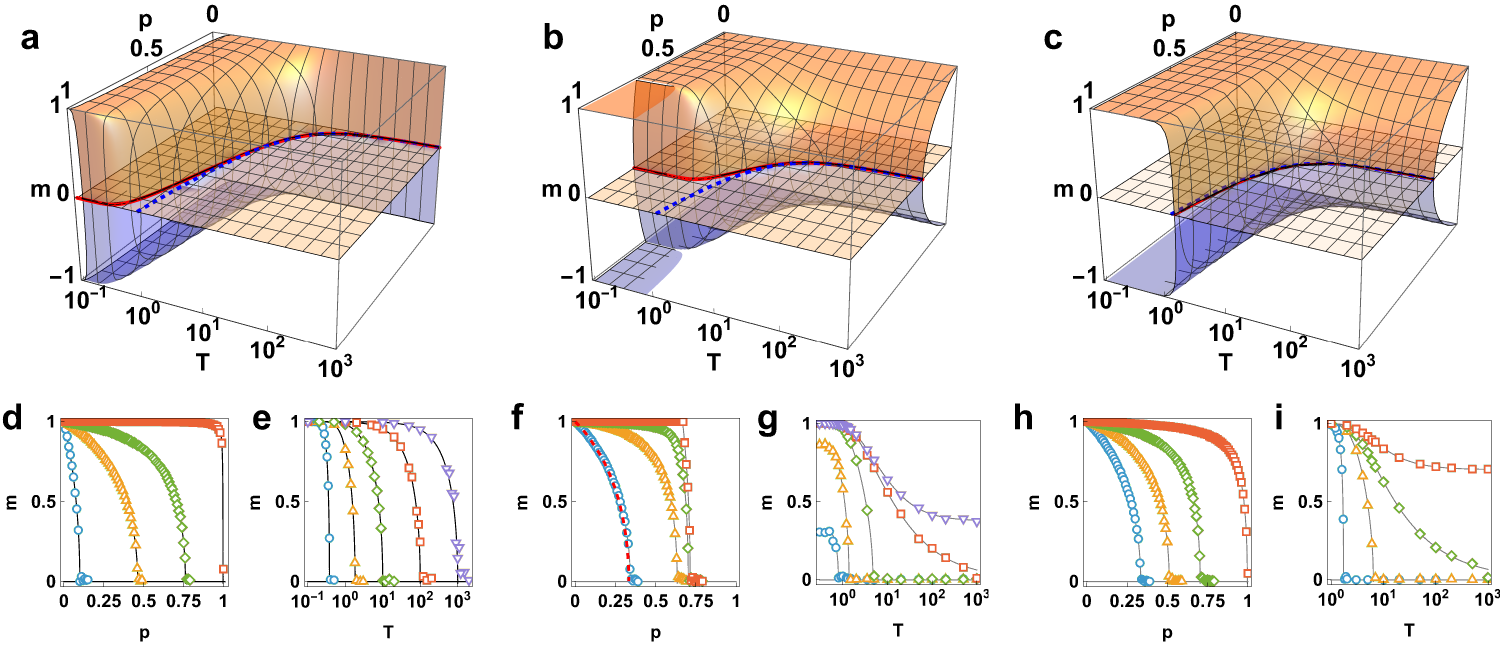}
\caption{Top row: solutions of the rate equation (\ref{eq:rate}) for (a) $q=1$, (b) $q=2$ and (c) $q=3$. The red solid line in the $m=0$ plane is $p^*_q(T)$ given by (\ref{eq:sep1}), while dashed lines come from the Taylor expansion thereof given by (\ref{eq:pq}). Bottom row: comparison of numerical simulations and analytical predictions of $m(p)$ and $m(T)$ for (d,e) $q=1$, (f,g) $q=2$ and (h,i) $q=3$. Solid lines are solutions of $F_1=0$, $F_2=0$ and $F_3=0$ and symbols are Monte Carlo simulations for (d) $T=10$ -- circles, $T=2$  -- triangles, $T=1$ -- diamonds, and $T=0.3$ -- squares, (e) $p=0.99$ -- circles, $p=0.5$ -- upward triangles, $p=0.1$ --diamonds, $p=0.01$ -- squares, and $p=0.001$ -- downward triangles, (f) $T=100$ -- circles, $T=2$ -- triangles, $T=1$ -- diamonds, $T=0.2$ -- squares, (g) $p=0.71$ -- circles, $p=0.68$ -- upward triangles, $p=0.5$ --diamonds, $p=1/3$ -- squares, and $p=0.3$ -- downward triangles; (h) $T=1000$ -- circles, $T=6$ -- triangles, $T=3$ -- diamonds, $T=1.73$ -- squares; (i) $p=1$ -- circles, $p=1/2$ -- upward triangles, $p=1/3$ --diamonds, $p=0.2$ -- squares. All simulations were performed for $N=2\cdot10^5$ nodes and $M=2\cdot10^5$ large MC steps \cite{data}.}
\label{fig:q123}
\end{figure*}

\begin{figure*}[!ht]
         \includegraphics[width=\textwidth]{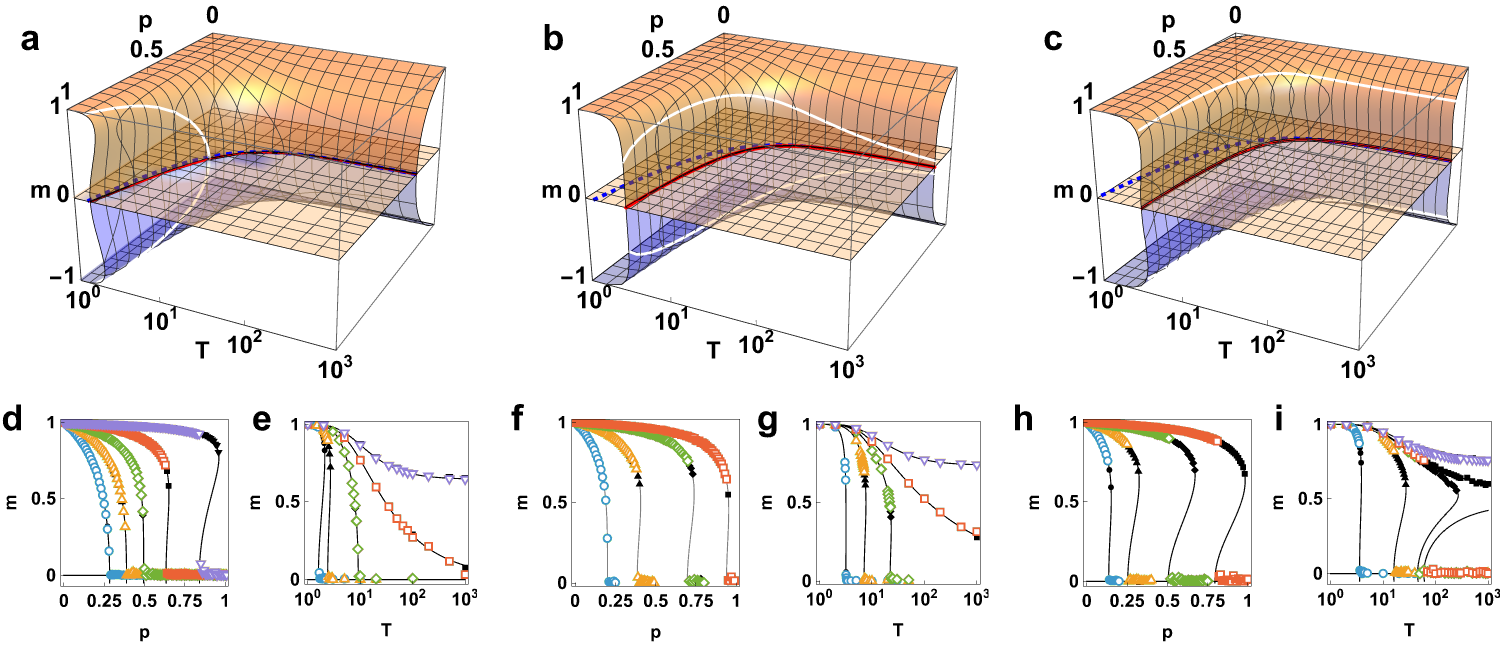}
\caption{Top row: solutions of the rate equation (\ref{eq:rate}) for (a) $q=4$, (b) $q=5$ and (c) $q=6$. The red solid line in the $m=0$ plane is $p^*_q(T)$ given by (\ref{eq:sep1}), while dashed lines come from the Taylor expansion thereof given by (\ref{eq:pq}). White solid lines denote the value of magnetization at the critical line $m(p^*_q(T))$.  Bottom row: comparison of numerical simulations and analytical predictions of $m(p)$ and $m(T)$ for (d,e) $q=4$, (f,g) $q=5$, and (h,i) $q=6$. Solid lines are solutions of $F_4=0$, $F_5=0$ and $F_6=0$ and symbols are Monte Carlo simulations for (d) $T=100$ -- circles, $T=10$ -- up triangles, $T=5.5$ -- diamonds, $T=3.5$ -- squares, and $T=2.2$ -- down triangles, (e) $p=1$ -- circles, $p=0.8$ -- up triangles, $p=0.4$ -- diamonds, $p=3/11$ -- squares, and $p=0.2$ -- down triangles, (f) $T=10^3$ -- circles, $T=7.5$ -- triangles, $T=4$ -- diamonds, $T=3.3$ -- squares, (g) $p=1$ -- circles, $p=0.4$ -- up triangles, $p=0.25$ -- diamonds, $p=0.2$ -- squares, $p=0.15$ -- down triangles; (h) $T=10^3$ -- circles, $T=10$ -- triangles, $T=5$ -- diamonds, $T=3.9$ -- squares; (i) $p=1$ -- circles, $p=0.2$ -- up triangles, $p=0.155$ -- diamonds, $p=0.15$ -- squares, $p=0.13$ -- down triangles. All simulations were performed for $N=2\cdot10^5$ nodes and $M=2\cdot10^5$ large MC steps \cite{data}. Full symbols denote ordered initial conditions ($\langle m \rangle=1$), while empty symbols -- random initial conditions ($\langle m \rangle=0$)}
\label{fig:q456}
\end{figure*}

\begin{table}[!ht]
\caption{Properties of phase transitions in the monolayer q-voter and q-Ising models, duplex versions thereof, and the q-Ising q-voter duplex examined in this study together with coefficients $a_q$ and $b_q$ of linear expansion (\ref{eq:pq}). The $C$ symbol stands for the continuous phase transition and $DC$ for a discontinuous one, while the dash "--" marks no transition.}\begin{tabular}{c|p{.75cm}p{.75cm}p{.75cm}p{.8cm}p{.8cm}p{.75cm}}
\hline\hline
     lobby size $q$ &  1 & 2 & 3 & 4 & 5 & 6\\
\hline
     q-Ising \cite{Jedrzejewski2015} & -- & -- & C & DC & DC & DC\\ 
     q-voter \cite{Nyczka2012} & -- & C & C & C & C & DC\\ 
     q-Ising duplex \cite{Chmiel2017} & C & C & C & C & C & C\\ 
     q-voter duplex \cite{Chmiel2015} & C & C & C & C & DC & DC\\ 
     q-voter and q-Ising & C & C & C & DC/C & DC/C & DC\\
     $a_q$ coefficient & 0 & $\frac{1}{3}$ & $\frac{1}{3}$ & $\frac{3}{11}$ & $\frac{1}{5}$ & $\frac{5}{37}$\\     
     $b_q$ coefficient & 1 & $\frac{8}{9}$ & 1 & $\frac{128}{121}$ & 1 & $\frac{1152}{1369}$\\
\hline\hline
\end{tabular}
\label{tab:prop}
\end{table}

\section{\texorpdfstring{Results for $q\le3$}{Results for q<=3}}
All the systems characterized by $q\le3$ share a common property: in their base (monolayer) versions there is either only a disordered state or a continuous phase transition between the disordered and order state (see Table \ref{tab:prop}). Therefore we are allowed to expect that their coupled version should exhibit similar properties. 

\subsection{\texorpdfstring{The $q=1$ case}{The q=1 case}}
We start with the $q=1$ case, for which the rate equation (\ref{eq:rate}) takes the form of a cubic equation that can be factorized as
\begin{equation}
F_1=m(B_1m^2+A_1)=0    
\end{equation}
with $B_1=\frac{1}{4}(p-1)(1-\mathrm{e}^{-\frac{2}{T}})$ and $A_1=\frac{1}{4}[(p-1)-(p+1)\mathrm{e}^{-\frac{2}{T}}]$
characterized by solutions that, after some rudimentary algebra, can be presented as
\begin{equation}
\left\{
\begin{aligned}
m_0 ~ ~ &= ~ 0\\
m_{1,2} &= \pm \sqrt{\frac{1-p\coth T^{-1}}{1-p}}
\label{eq:q1mag}
\end{aligned}\right.
\end{equation}
Given the form of $m_0,m_1$ and $m_2$, the system undergoes continuous phase transition between the ordered phase and the disordered one both when the control parameter is $T$ and $p$ as shown in Fig.~\ref{fig:q123}a and confirmed by performing extensive Monte Carlo simulations (Fig.~\ref{fig:q123}d,e). Seemingly unexpected, as both base models exhibit no transitions, this phenomenon is already present in the q-Ising and q-voter duplex case (Table \ref{tab:prop}). Following, we arrive at a particularly concise relation for the critical line, i.e.,
\begin{equation}
    p^*_1(T) = \tanh T^{-1},
\label{eq:pq1}
\end{equation}
shown in Fig.~\ref{fig:q123}a with solid line. The linear expansion takes an exceptionally simple form ($a_1=0$, $b_1=1$) that can be written out as 
\begin{equation}
    pT = 1,
\label{eq:pTq1}
\end{equation}
presented with a dashed line in Fig. \ref{fig:q123}a. It follows that for $T \gg 1$ noise in the coupled model with $q=1$ can be treated as an \textit{opposite measure to temperature}, i.e., for a given $T$ one has to secure at least $p=1/T$ noise probability to prevent the system moving from a disordered state to an ordered one. This result comes as a second key point raised in this study.

\subsection{\texorpdfstring{The $q=2$ case}{The q=2 case}}
Let us now examine the $q=2$ case, for which (\ref{eq:rate}) takes the form of a quintic equation that can be factorized as a product of $m$ and a biquadratic equation:
\begin{equation}
    F_2 = m(C_2m^4+B_2m^2+A_2)
\label{eq:f2}
\end{equation}
with $C_2=(p-1) \left(1-e^{-4/T}\right)$, $B_2=(4 p-2) e^{-4/T}-8 p+6$ and $A_2=(5 p+1) e^{-4/T}+7 p-5$ and an obvious set of solutions
\begin{equation}
\left\{
\begin{aligned}
m_0 ~  &= ~ 0\\
m_{1,2} &= \pm \sqrt{\frac{B_2 +\sqrt{B_2^2-4A_2C_2}}{2C_2}}\\
m_{3,4} &= \pm \sqrt{\frac{B_2 -\sqrt{B_2^2-4A_2C_2}}{2C_2}}
\label{eq:q2mag}
\end{aligned}\right.
\end{equation}
The last two solutions either bring no relevant values ($|m_{3,4}|>1$ or they are equal to $\pm 1$, but in such a case, they are unstable (see Appendix \ref{app1} for details). As for $q=1$, also here, the system exhibits a continuous phase transition for both control parameters (see analytical solutions of Eq.~(\ref{eq:f2}) in Fig.~\ref{fig:q123}b and Monte Carlo simulations in Fig.~\ref{fig:q123}f and in Fig.~\ref{fig:q123}g). The relation between the critical value of $p$ and the temperature can once again be obtained using Eq.~(\ref{eq:sep1}) or more directly by solving $A_2=0$ resulting in
\begin{equation}
p^{*}_2(T) = \frac{5-\mathrm{e}^{-\frac{4}{T}}}{7+5\mathrm{e}^{-\frac{4}{T}}},    
\label{eq:pq2}
\end{equation}
and is marked with the solid line in Fig.~\ref{fig:q123}b. Unlike $q=1$, one can identify ranges of $p$ for which the system stays strictly in an ordered or disordered state, regardless of the temperature. The limiting values are found by exploring the limits $T \rightarrow \infty$ and $T \rightarrow 0$ in Eq.~(\ref{eq:pq2}). The first limit trivially coincides with the monolayer q-voter model with $q=2$ \cite{Nyczka2012} (dashed line in Fig.~\ref{fig:q123}b), the second one $p^*_2(T \rightarrow 0)=5/7$ is unexpected and cannot be explained by relating to the properties of either of the base models. Even for low temperatures, the system still stays in the disordered state once $p>5/7$. This phenomenon can be spotted by examining $F_2=0$ for $T \rightarrow 0$ in which case it takes the form of $m = \pm 1$ for $p \le 2/3$, $m = \pm \sqrt{(5-7p)/(1-p)}$ for $p \in (2/3;5/7]$ and $m=0$ for $p \in [5/7,1]$. These unexpected predictions are fully confirmed by Monte Carlo simulations shown in Fig.~\ref{fig:q123}f. Let us mention here that the $q=2$ case also brings surprising results in the context of the partially duplex clique model \cite{Chmiel2017}.

\subsection{\texorpdfstring{The $q=3$ case}{The q=3 case}}
The $q=3$ case, shown in Fig. \ref{fig:q123}c comes (like the previous ones) with continuous-only phase transitions (see Appendix \ref{app2} for the full form of $F_3$). The critical line calculated using Eq.~(\ref{eq:sep1}) reads  
\begin{equation}
    p_3^{*}(T)=\frac{14+3 e^{-2/T}- e^{-6/T}}{3 \left(2+9 e^{-2/T}+5 e^{-6/T}\right)}
\end{equation}
and can be approximated by $p^*_3(T) = 1/3+1/T$ for $T \gg1$, closely covering the exact value (see solid and dashed lines in Fig.~\ref{fig:q123}c). A notable difference between $q=3$ and $=1,2$ is connected to the fact that in addition to recovering the monolayer q-voter behavior for $ T\rightarrow\infty$ (see \ref{fig:q123}h) when $p=1$, one should arrive at the original results of the monolayer q-Ising model. Below the critical temperature $T_c \approx 1.7214$ \cite{Jedrzejewski2015,Chmiel2024}, one observes no disordered state regardless of the value of $p$.

\section{\texorpdfstring{Results for $q\ge 4$}{Results for q>=4}}

The fundamental difference between the previously described versions with $q < 4$ and the ones with $q \ge 4$ is that while the former are characterized by solely continuous phase transitions, the building blocks of the latter exhibit a discontinuous one, accompanied by a hysteresis. However, due to the fact that the width of the hysteresis changes non-monotonically with $q$ for the q-Ising model \cite{Jedrzejewski2015} and that the q-voter model exhibits discontinuous phase transitions for $q \ge 6$ as compared to the q-Ising one, where it happens already for $q \ge 4$ (see Table \ref{tab:prop} for an overview), coupling brings varied results for different $q$. For the sake of clarity of presentation, we further refrain from giving full forms of relevant rate equations and critical lines and present them in detail in Appendix \ref{app2}.

\begin{figure*}[!ht]
         \includegraphics[width=\textwidth]{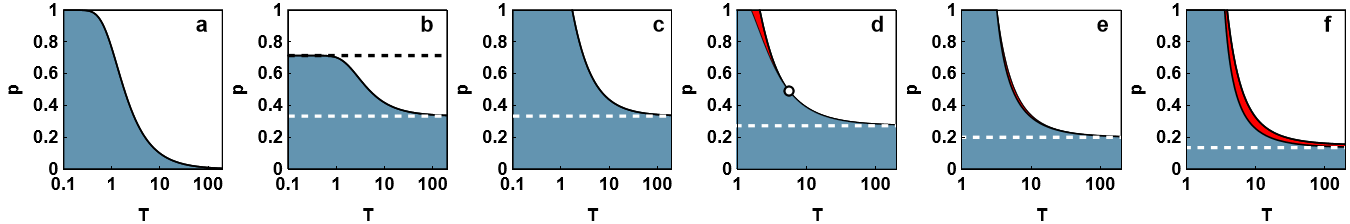}
\caption{Phase diagrams for (a) $q=1$, (b) $q=2$, (c) $q=3$, (d) $q=4$, (e) $q=5$, (f) $q=6$. Blue regions denote the ordered phase, white regions -- the disordered one, while red marks the region of bistability. White dashed lines in panels (b-f) mark critical values of $p$ for the relevant monolayer q-voter model, i.e., $p^*_q(T\to\infty)$, while the black dashed line in panel (b) reflects $p^*_2(T\to0)$. The single data point on panel (d) denotes the tricritical point for $q=4$ given by (\ref{eq:tricri}). Note the different scale on the x-axis in panels (a-c) and (d-f).}
\label{fig:diag}
\end{figure*}

\subsection{\texorpdfstring{The $q=4$ case}{The q=4 case}}
The $q=4$ case is of particular interest in the context of this study due to the fact that although the monolayer q-voter model is characterized by a continuous phase transition with $p^*_4=3/11$, in the q-Ising model we originally observe two spinodals: lower for $T^{(1)}_c \approx1.5886$ and upper for $T^{(2)}_c \approx 2.1092$, indicating a region of bistability in the phase diagram. This in turn suggests that while at $p=1$ we need to recover the discontinuous transition, we should observe a tricritical point for a specific value of $p_{\rm{tri}}$, with a coexistence of the disordered, ordered, and bistable phases. 

This assumption is confirmed in Fig. \ref{fig:q456}a, where apart from the numerical solution of $F_4=0$ we also show the magnetization calculated for the points lying on the critical line $p^*_4(T)$ obtained from Eq.~(\ref{eq:sep1}).

To obtain the coordinates of the tricritical point $(T_{\rm{tri}},p_{\rm{tri}})$ we follow Landau description of equilibrium phase transitions, calculating the effective potential

\begin{equation}
V_q(m)=-\int F_q(m)dm    
\end{equation}
for $q=4$ and expanding it into the power series, keeping the first three terms \cite{Nyczka2012,Jedrzejewski2019}
\begin{equation}
V_4(m)=A^L_4(T,p)m^2+B^L_4(T,p)m^4+C^L_4(T,p)m^6.    
\end{equation}
The value of $T$ for which $B^L_4(T,p^*_4(T))$ changes its sign (provided that $C^L_4(T,p)>0$), indicates the tricrtical point that can be extracted as the only real, non-negative solution of $-55 u^4-72 u^3+358 u^2+40 u-15=0$ where $u=\exp(4/T_{\rm{tri}})$. Combined with $p^*_4(T_{\rm{tri}})$ leads to following coordinates of the tricritical point
\begin{equation}
(T_{\rm{tri}},p_{\rm{tri}}) \approx (5.6331, 0.4888),
\label{eq:tricri}
\end{equation}
shown in Fig. \ref{fig:diag}d. The importance of this result comes from the fact that it marks the areas of influence of the q-Ising and q-voter model --- below $T_{\rm{tri}}$ the q-Ising model is sufficiently ``strong'' to make the coupled system exhibit a discontinuous phase transition, while above we have only continuous transitions, i.e., the q-voter dynamics prevails. Indirectly, it can also be understood as the point of balance between temperature-like and noise-like dynamics. 

\subsection{\texorpdfstring{The $q=5$ case}{The q=5 case}}
Judging from the properties gathered in Table~\ref{tab:prop}, one expects that the coupled $q=5$ system should possess similar properties to the $q=4$ one. However, the system with $q=5$ can be characterized by a finite value of magnetization at the critical line even for large $T$ (cf. Fig.~\ref{fig:q456}b,d,g, where we numerically calculated the magnetization for $m(T,p^*_5(T))$ along with numerical simulations). This result is also confirmed in the Landau approach  (see Appendix \ref{app3}) by exploring $B(T,p)$ that approaches zero as $-T^{-1}$ for $T \gg 1$. Apparently, $m(T,p^*_5(T))$ is not a monotonic function and takes a maximum for $(T \approx 4.3207, p \approx 0.6309)$. As a result, we observe a narrow strip of bistability along the critical line that non-monotonically changes its width. These interesting differences between $q=4$ and $q=5$ can be attributed to the character of the phase transition for the base q-voter, i.e., a tricritical phase transition \cite{Jedrzejewski2019} where the magnetization at the critical point scales as $m \sim (p_c-p)^{\beta}$ with $\beta=1/4$ (unlike $q<5$ where $\beta=1/2$).

\subsection{\texorpdfstring{The $q=6$ case}{The q=6 case}}
The last example in a set of systematically examined systems is the one with $q=6$. Here, in line with Table~\ref{tab:prop}, we observe only discontinuous phase transitions (cf. Fig.~\ref{fig:q456}c), although there is a maximum of the magnetization calculated at the critical line $p^*_6(T)$. If $p$ stays between the value of the lower and upper spinodal of the base q-voter model, the system remains in the bistable state for almost the whole range of $T$ (see Fig.~\ref{fig:q456}i). 

The results for all the examined values of $q$ are presented as phase diagrams in Fig.~\ref{fig:diag}, comprehensively gathering types of exhibited phase transitions and the range of parameters connected to them.

\begin{figure}[!ht]
         \includegraphics[width=\columnwidth]{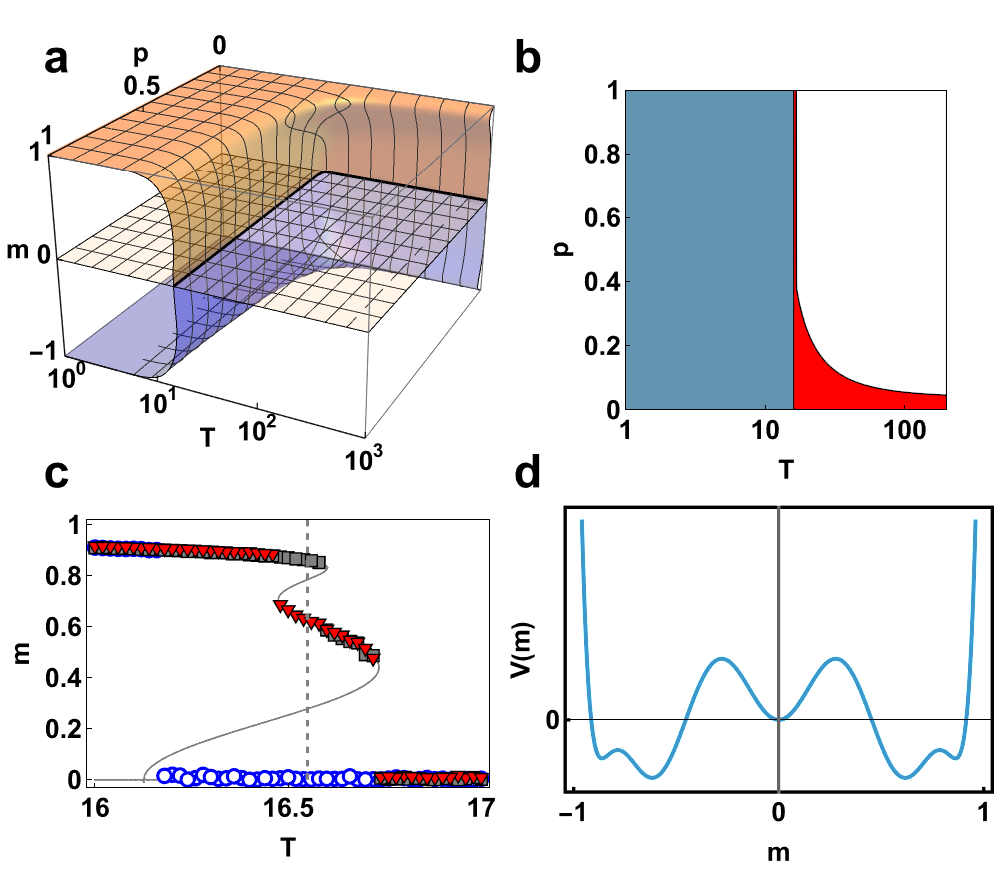}
\caption{Results for $q=20$. (a) Solution for the rate equation $F_{20}=0$ (b) Phase diagram for $q=20$. (c) Comparison of analytical results for $q=20$ and $p=0.39$ (solid line) and numerical simulations (symbols) \cite{data}. Circles denote simulations starting from the disordered initial conditions $\langle m \rangle = 0$, squares - the ones starting from the fully ordered state $\langle m \rangle = 1$, and triangles - from $\langle m \rangle = 0.4$. The dashed line marks the value of $T$ for which the effective potential was calculated, see panel (d). All simulations were performed for $N=10^6$ nodes and $M=2\cdot10^5$ large MC steps. (d) Effective potential $V_{20}(m)$ for $p=0.39$ and $T=16.55$ }
\label{fig:q20}
\end{figure}

\section{\texorpdfstring{Higher values of $q$}{Higher values of q}}
In the two previous sections, we have presented the results for $q=1,..,6$ in detail, backed by explicit formulas given in Appendix \ref{app2}. We refrain from performing similar calculations for $q \ge 7$ due to the fact that $q=6$ marks the point for which both base models exhibit only discontinuous phase transitions. This in turn allows us to tacitly assume that the phase diagram of $q=7$ coupled model should largely resemble the one presented in Fig.~\ref{fig:diag}f.  

However, when considering the behavior of the coupled system for larger values of $q$, we note that in the monolayer q-Ising both the lower spinodal and the width of the hysteresis grow linearly with $q$ \cite{Jedrzejewski2015}, i.e., we should expect the ordered phase to move towards larger values of $T$ on the phase diagram and coexistence phase to increase its width for $p$ close to 1. On the other hand, for $q \gg 1$ the lower spinodal in monolayer q-voter model reads $2q/2^{q}$, and the upper one also decreases with $q$ \cite{Nyczka2012}. It follows that for large values of $q$ the phase diagram will be mainly governed by the q-Ising dynamics, only for small values of $p$ the q-voter dynamics shall prevail. Figure \ref{fig:q20} presents the results for $q=20$: we see the $p^*_{20}(T)$ is almost perpendicular to $T$ axis, however the area of phase coexistence is much larger (cf. Fig.~\ref{fig:q20}a,b) than for any of the previously analyzed cases in Fig.~\ref{fig:diag}. Interestingly, owing to the fact that both discontinuous transitions mask each other, in a narrow strip of the parameters (cf. Fig.~\ref{fig:q20}c) we encounter a successive phase transition \cite{Mutsuki2019} of two consecutive discontinuous transitions similar to the one observed in an asymmetric duplex q-voter model \cite{Chmiel2020}. This observation is confirmed while examining the effective potential $V_{20}(m)$ in Fig.~\ref{fig:q20}d that possesses five minima, reflecting stable solutions of $F_{20}$.

Comparison of Fig. \ref{fig:diag}f and Fig. \ref{fig:q20}b suggests that for higher $q$ the q-Ising dynamics should dominate the phase diagram. In particular, Fig.~\ref{fig:q20}a clearly shows that results for $p > 1/2$ are identical or close to the ones obtained for $p=1$, i.e., for the base q-Ising model.       

\begin{figure}[!ht]
         \includegraphics[width=\columnwidth]{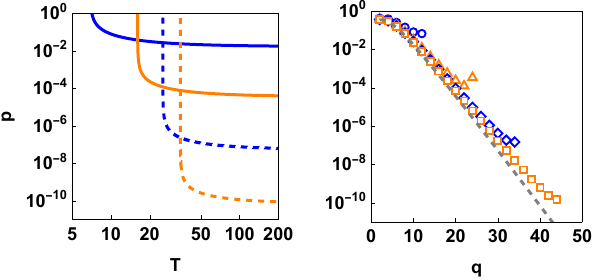}
\caption{Higher values of $q$. (a) Log-log plot of the critical line for $p^*_q(T)$ given by Eq.~(\ref{eq:apppq}) for $q=10$ (dark solid line), $q=20$ (light solid line), $q=30$ (dark dashed line), and $q=40$ (light dashed line), (b) Semi-log plot of critical values $p^*_q(T)$ versus $q$ given by Eq.~(\ref{eq:apppq}) for $T=10$ (circles), $T=20$ (triangles), $T=30$ (diamonds), and $T=40$ (squares).}
\label{fig:fig6}
\end{figure}

To examine the character of the lower spinodal in a more quantitative way, we derived a fully analytical formula for the critical line $p^*_q(T)$, shown separately as Eq.~(\ref{eq:apppq}) in Appendix \ref{app4} due to its complicated form. This allows us to show in Fig.~\ref{fig:fig6}a that even for relatively small values of lobby size, the critical line runs almost vertically between $p=1$ and negligible values of $p$ at which it then stays regardless of the temperature. Although the limiting behavior for $q \to \infty$ is not traceable with Eq.~(\ref{eq:apppq}) due to the mentioned linear relation between the critical temperature in the monolayer Ising model and the lobby size, we can still spot how the coupled system behaves at given $T$ with increasing $T$. Figure~\ref{fig:fig6}b suggests that the limiting case $p^*_q(T \to \infty) = 2(q-1)/[(2(q-1)+2^q]$ which coincides with the monolayer q-voter lower spinodal (see Appendix \ref{app4}) serves as a satisfactory approximation.

\section{Discussion}
In summary, we have presented the effects of an interplay between two opinion dynamics models with limited influence of neighbors characterized by, respectively, temperature-like and noise parameters. 

Let us stress that there have been several works that tackle the issue of different types of dynamics taking place in the multilevel network. Probably the most prominent example is the coupled system of two spreading processes: awareness and epidemic \cite{Granell2013}. This study was followed by other works coupling threshold dynamics  \cite{Czaplicka2016}, traffic rules \cite{Chen2020} or even the q-voter model \cite{Jankowski2022} with simple epidemics (SIS or SIR) that have proven to be an important modeling tool in the face of so-called infodemics (epidemics of information) \cite{Gradon2021}.

However, in our study, we set the focus on comparing two types of dynamics mimicking the phenomenon of changing one's opinion under the influence (pressure) of a group of unanimous neighbors. Although the simulated social process is in general the same in both layers, there is a difference in parametrization (temperature-like parameter $T$ in the case of the q-Ising model and probability $p$ of an independent choice in the q-voter one) that allows judging the relation between $T$ and $p$.

The properties of the base models (no phase transitions, continuous and discontinuous transitions between the ordered and disordered phases) have allowed us to expect a rich behavior in the coupled system. Nonetheless, as clearly shown in Fig. \ref{fig:diag}, each of the first six values of $q$ presents a strikingly different picture, characterized by different properties -- some of them unexpected, as in the case of $q=2$. This fact emphasises that the size of the lobby plays an important role.  

The very construction of the coupled model gives the possibility to obtain the exact relation for the critical line $p^*_q(T)$, separating ordered and disordered states, virtually for any value of $q$, which takes a particularly simple form of $p^*_q(T)=\tanh T^{-1}$ if the lobby consists simple of one randomly chosen neighbor. Moreover, we also showed that in the high temperature limit, $p^*_q(T)$ can be rewritten as $T(p-a)=b$ with $a$ being the critical value of $p$ for the base q-voter model. Although only holding for the critical line, this relation suggests that temperature can be treated as a measure opposite to noise represented here by the independence probability $p$. This observation is most prominent for $q=1$, where in the high temperature limit we obtain $pT=1$.

Let us also bring to the front the fact that the coupling can be used to determine the range of influence of the given type of dynamics over the other one. This phenomenon is visible for $q=4$, where, based on Landau theory, we calculated the coordinates of the tricritical point; clearly for $T>T_{\rm{tri}}$ we have only continuous phase transitions connected to the q-voter model, while for $T<T_{\rm{tri}}$ only discontinuous transition takes place, that is the landmark of the q-Ising model for such $q$. However, as the $q$ grows, the q-Ising dynamics seems to prevail, as shown for $q=20$ in Fig.~\ref{fig:q20}. A possible explanation is connected to differences in the rules of the base models: while the q-voter requires the same opinion among the members of the lobby, the q-Ising dynamics is not that strict. 

As mentioned in Sec. \ref{sec:model}, we have chosen the type of coupling (AND dynamics) motivated by its widespread use in the previous studies, providing an opportunity for a direct comparison. However, it is also possible to consider OR coupling, when the change is accepted if either of dynamics suggests a flip. From the user-centered point of view connected to multiplex networks, this rule brings to the front the fact that the opinion can be changed even if only one community (i.e., one layer) prompts for it. In such a situation the probability of changing a down spin to an up one is proportional to $\gamma^{+}_I(1-\gamma^{+}_v)+\gamma^{+}_v(1-\gamma^{+}_I)+\gamma^{+}_I \gamma^{+}_v$ (the spin flips if the q-Ising dynamics suggests a change and the q-voter one does not or if the q-voter dynamics suggests a change and the q-Ising one does not or if both do so). An opposite situation comes with $\gamma^{-}_I(1-\gamma^{-}_v)+\gamma^{-}_v(1-\gamma^{-}_I)+\gamma^{-}_I \gamma^{-}_v$ and as a result, the rate equation for the OR case reads

\begin{equation}
    F_q^{OR} = F^{I}_q+F^{v}_q-F^{AND}_q
\label{eq:rateor}
\end{equation}
where $F^{I}_q=\frac{1}{2}[(1-m)\gamma^{+}_I-(1+m)\gamma^{-}_I]$ is the rate equation for the monolayer q-Ising model, $F^{v}_q=\frac{1}{2}[(1-m)\gamma^{+}_v-(1+m)\gamma^{-}_v]$ is the equivalent of the q-voter dynamics and $F^{AND}_q \equiv F_q$ given by Eq.~(\ref{eq:rate}) is the rate equation for the AND version examined in this study. The rate equation (\ref{eq:rateor}) proves that although OR dynamics is connected to the AND one, the base models play an important role that might suppress the coupling effect. Indeed, for the simplest case ($q=1$) one can easily obtain from (\ref{eq:rateor}) that $m_0=0$ and $m_{1,2}=\sqrt{(p+1)-(p-3) e^{-2/T}}/\sqrt{(1-p) \left(1-e^{-2/T}\right)} > 1$, i.e., the system stays in the disordered state regardless of $p$ and $T$. This initial result is in line with the monolayer behavior of q-voter or q-voter dynamics for $q=1$ (see Table \ref{tab:prop}) while largely different than the system described by $F_1=0$ (see Fig. \ref{fig:q123}a) and also consistent with the lack of reinforcement introduced by the AND rule. However, these considerations are merely a glimpse into the problem of the OR dynamics that calls for a separate study that could examine it in detail.

In the same way, the presented model suffers from two obvious oversimplifications: the first one is connected to the limited number of opinions, while the second one is linked to the topology. Reaching outside the binary setting has solid psychological grounds, e.g., it gives the possibility to introduce an intermediate opinion ($s_i=0$) that could denote an unconvinced state. While the q-voter model has in fact been generalized for a higher number of states \cite{Nowak2022,Lipiecki2022}, one would first need to find the equivalent of q-Ising version of the multi-state Ising model, i.e., the Potts model \cite{Potts1952}. Additionally, as can be seen for the multi-state q-voter model \cite{Nowak2022,Lipiecki2022}, even a simple shift to a three-state dynamics opens the field for different ways of accounting, e.g., for the anticonformity (e.g., taking any opinion other than the q-lobby \cite{Nowak2022} versus moving away from the q-lobby opinion \cite{Lipiecki2022}. Therefore, although a multi-state version of the problem considered in this study might be psychologically relevant, it cannot be achieved in a straightforward way.

As it concerns the underlying topology, the coupled model could be examined in a more real-world-like topologies, such as a heterogeneous networks with scale-free degree distribution $p(k)\sim k^{-\gamma}$, both numerically and analytically, using the pair-approximation (PA) approach introduced in the case of the q-voter in \cite{Jedrzejewski2017}. However, as shown by Jedrzejewski \cite{Jedrzejewski2017} and  Gradowski and Krawiecki \cite{Gradowski2020}, the results for large average degree $\langle k \rangle$ overlap with the mean-field approach for the complete graph, for small $\langle k \rangle$ shifting the transition point close to $p=0$, On the other hand, Krawiecki and Gradowski have shown \cite{Krawiecki2023} that the introduction of heterogeneity to the q-Ising duplex with partial overlap \cite{Chmiel2017} might change the character of the phase transition in specific cases. Therefore, making predictions on the behavior of our coupled system under large heterogeneity is difficult beforehand.

Finally, we believe that our results can shed new light on the discussion on the understanding of the concept of social temperature in opinion dynamics models.

\section{Acknowledgements}
This research was funded by POB Research Centre Cybersecurity and Data Science of Warsaw University of Technology, Poland within the Excellence Initiative Program—Research University (ID-UB).

\appendix
\section{\texorpdfstring{Limiting behavior $T\to0$ for $q=2$}{Appendix A}}\label{app1}

\begin{figure}[!ht]
         \includegraphics[width=\columnwidth]{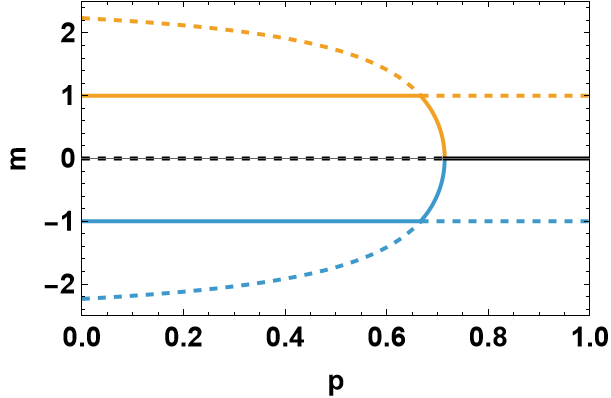}
\caption{Magnetization $m$ versus noise $p$ for $q=2$ and $T\to0$ given by Eq.~(\ref{eq:q2T0}). Solid lines are stable solutions and dashed ones -- unstable.}
\label{fig:q2T0}
\end{figure}

\begin{figure}[!ht]
         \includegraphics[width=\columnwidth]{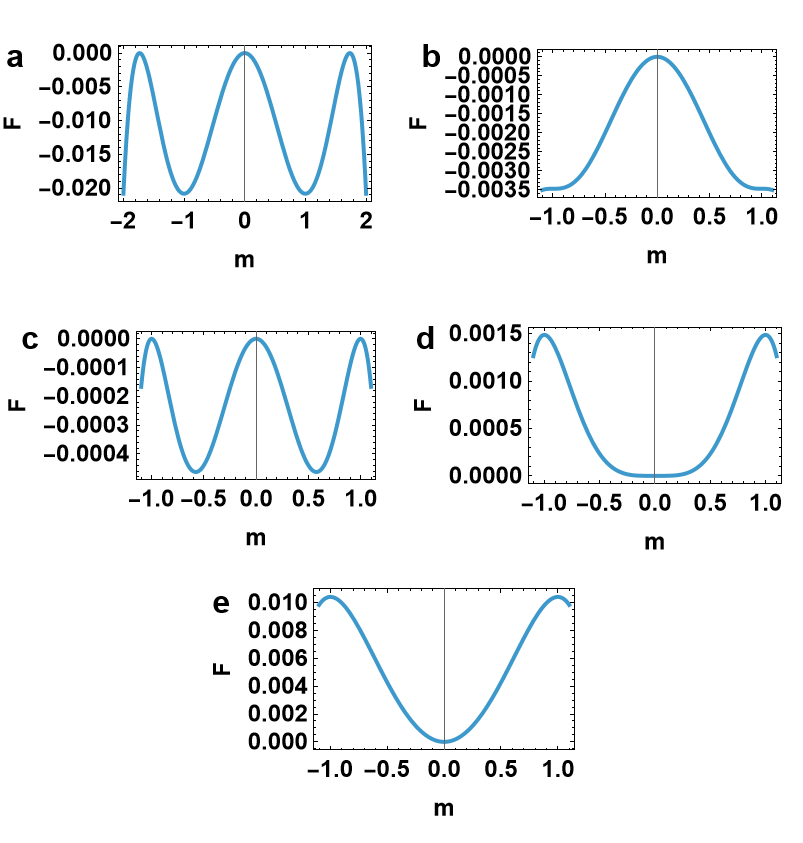}
\caption{Effective potential $V_2(m)$ for (a) $p=0.5$, (b) $p=2/3$, (c) $p=0.7$, (d) $p=5/7$, and (e) $p=0.8$.}
\label{fig:q2V}
\end{figure}

Let us note that the rate equation for $q=2$ can be expressed as 
\begin{equation}
F_2(m) = m \left(1-m^2\right) \left(m^2 (p-1)-7 p+5\right)
\end{equation}
when $T\to0$. The five solutions
\begin{equation}
\left\{.
\begin{aligned}   
m_0 &= 0\\
m_{1,2} &= \pm\sqrt{\frac{5-7p}{1-p}}\\
m_{3,4} &= \pm1
\end{aligned}\right.
\label{eq:q2T0}
\end{equation}
are presented in Fig.~\ref{fig:q2T0} 
where the stability of solutions has been resolved using the effective potential $V_2(m)=-\int F_2dm $ (see Fig. \ref{fig:q2V}). The analysis proves that the solutions $m=\pm1$ seen in Fig. 2b in the main text for $p>5/7$ are unstable. Additionally, we also see that the solution $m=\pm1$ loses stability in favor of $m=\pm\sqrt{(5-7p)/(1-p)}$ once $p>2/3$.

\section{\texorpdfstring{Explicit formulas for $3 \le q \le 6$ and the limiting behavior for $T\to\infty$ and $p=1$}{Appendix B}}\label{app2}

Using the general rate equation (\ref{eq:rate}) for the coupled system, we obtain the following explicit formula for $q=3$:

\begin{equation}
    F_3 = m\left(A_3+B_3m^2+C_3m^4+D_3m^6\right)=0
\end{equation}
with 
\begin{equation}
\begin{aligned}   
A_3 &= -(15 p+1) e^{-6/T}+3 (1-9p)e^{-2/T}+2 (7-3 p)\\
B_3 &= (3-19 p) e^{-6/T}+3 (11 p-3) e^{-2/T}+2(p-5)\\
C_3 &= (1-p)(-3 e^{-6/T}+9 e^{-2/T}-6)\\
D_3 &=(1-p) (e^{-6/T}-3e^{-2/T}+2) 
\end{aligned}
\end{equation}
and the following equation for the critical line
\begin{equation}
    p_3^{*}(T)=\frac{14+3 e^{-2/T}- e^{-6/T}}{3 \left(2+9 e^{-2/T}+5 e^{-6/T}\right)} \underset{T \to \infty}= \frac{1}{3}
\end{equation}

The equation shows the critical value of noise for the base q-voter model. Requiring that $p_3^*(T_c)=1$ leads to the following equation $1-2u^3-3u=0$, where $u=e^{-2/T_c}$ with only real solution $T=-2\ln^{-1}(r/2-1/r)$, where $r=[2(1+\sqrt{3})]^{1/3}$, i.e, $T_c \approx 1.7214$ -- the critical temperature for monolayer q-Ising model. Thus, we see that the coupled model correctly gives limiting behavior of the base q-voter and q-Ising models.\\ 

For $q=4$, we have
\begin{equation}
    F_4 = m\left(A_4+B_4m^2+C_4m^4+D_4m^6+E_4m^8\right)=0
\end{equation}
with
\begin{equation}
\begin{aligned}   
A_4 &= -(39 p+1)e^{-8/T}+(4-100 p) e^{-4/T}+(45-37 p)\\
B_4 &= (80 p-16) e^{-4/T}+(4-84 p) e^{-8/T}+4 (9 p-5)\\
C_4 &= -2 (p+3) e^{-8/T}+8 (p+3) e^{-4/T}+10 p-34\\
D_4 &= 4 (1-p) \left(-4 e^{-4/T}+3+e^{-8/T}\right)\\
E_4 &= (1-p) \left(4 e^{-4/T}-3-e^{-8/T}\right)
\end{aligned}
\end{equation}

and the following equation for the critical line
\begin{equation}
p_4^{*}(T)=\frac{45+4 e^{-4/T}-e^{-8/T}}{37+100 e^{-4/T}+39e^{-8/T}} \underset{T \to \infty}= \frac{3}{11}
\end{equation}
Here, similarly to $q=3$ case we can require $p_3^*(T_c)=1$ which leads to $1-5u^2-12u=0$, where $u=e^{-4/T_{1}}$. The only real solution is $T_1=-4/\log\left[\frac{1}{5} \left(\sqrt{41}-6\right)\right]\approx1.5886$, i.e., the value of the lower spinodal for the base q-Ising model.

Let us note that when $p=1$, higher terms ($D_4$ and $E_4$) disappear and we are left with a simple quintic equation that can be factorized as $F_4 = m\left(A_4+B_4m^2+C_4m^4\right)=0$  with obvious solutions:
\begin{equation}
\begin{aligned}
m_0 ~  &= ~ 0\\
m_{1,2} &= \pm \sqrt{\frac{B_4 +\sqrt{B_4^2-4A_4 C_4}}{2C_4}}\\
m_{3,4} &= \pm \sqrt{\frac{B_4 -\sqrt{B_4^2-4A_4 C_4}}{2C_4}}
\label{eq:quad}
\end{aligned}
\end{equation}

This allows us to calculate the value for the upper spinodal, making use of the fact that at this point $m_1$ and $m_3$ should be equal, which leads to $B_4^2-4A_4C_4=0$ and in turn to $5 u^4-8 u^3+10 u^2-8 u+1=0$ where $u=e^{-4/T_{2}}$. The only real, non-negative solution is then $T_2 \approx 2.1092$.\\

For $q=5$, we have
\begin{equation}
\begin{aligned}
    F_5 &= m\left(A_5+B_5m^2+C_5m^4+D_5m^6+E_5m^8+F_5m^{10}\right)\\
    &=0   
\end{aligned}
\label{eq:f5}
\end{equation}
with

\begin{equation}
\begin{aligned}
\begin{split}
A_5 &= (-95 p-1) e^{-10/T}+(5-325 p) e^{-6/T}+\\
&+(30-350 p) e^{-2/T}+130 p+94\\
B_5 &= 25 (p-1) e^{-6/T}+(750 p-110) e^{-2/T}+\\
&+(5-325 p) e^{-10/T}+5 (26-90 p)\\
C_5 &= -2 (43 p+5) e^{-10/T}+(270 p+50) e^{-6/T}+\\
&+(140-460 p) e^{-2/T}+404 p-308\\
D_5 &= 10 (1-p) \left(e^{-10/T}-5 e^{-6/T}-6 e^{-2/T}+10\right)\\
E_5 &= 5 (1-p) \left(-e^{-10/T}+5 e^{-6/T}-2 e^{-2/T}-2\right)\\
F_5 &= (1-p) \left(1-e^{-2/T}\right)^2 \left(e^{-6/T}+2 e^{-4/T}-2 e^{-2/T}-6\right)
\end{split}
\end{aligned}
\label{eq:f5c}
\end{equation}

and the following equation for the critical line
\begin{equation}
\begin{aligned}
p_5^{*}(T)&=\frac{94+30 e^{-2/T}+5 e^{-6/T}-e^{-10/T}}{5 \left(-26+70 e^{-2/T}+65 e^{-6/T}+19e^{-10/T}\right)}\\
&\underset{T \to \infty}= \frac{1}{5}
\end{aligned}
\label{eq:p5}
\end{equation}

As for $q=4$, the condition for the lower spinodal is $p_5^{*}(T_1)=1$ and leads to $-3 u^5-10 u^3-10 u+7 = 0$ where $u=e^{-2/T_{1}}$. The only real, non-negative solution is then $T_1 \approx 3.1913$. To calculate the upper spinodal we once again use the fact that for higher terms ($D_5$, $E_5$, and $F_5$) disappear for $p=1$ and thus we can utlize Eq.~(\ref{eq:quad}), changing $A_4$, $B_4$ and $C_4$ coefficients into $A_5$, $B_5$ and $C_5$ which leads to
\begin{equation}
4 u^{10}-15 u^6+20 u^5-10 u^3+1 = 0
\end{equation}
where $u=e^{-2/T_{2}}$, with only real, non-negative solution $T_2 \approx 3.2003$. Thus, both $q=4$ and $q=5$ monolayer q-Ising models are characterized by discontinuous phase transitions and hysteresis, however in the latter case the width the hysteresis is rather small (i.e., over 50 times smaller than for $q=4$).

Finally for $q=6$, we have
\begin{equation}
\begin{aligned}
\begin{split}
    F_6 &= m\left(A_6+B_6m^2+C_6m^4+D_6m^6+E_6m^8\right.\\
    &\left.+F_6m^{10}+G_6m^{12}\right)=0
\end{split}
\end{aligned}
\end{equation}
with
\begin{equation}
\begin{aligned}
\begin{split}
A_6 &= (-223 p-1) e^{-12/T}+(6-966 p) e^{-8/T}+\\
&+(45-1485 p) e^{-4/T}+306 p+270\\
B_6 &= -12 (77 p+3) e^{-8/T}+30 (87 p-7) e^{-4/T}+\\
&+(6-1126 p) e^{-12/T}-880 p+560\\
C_6 &= 18 (91 p+5) e^{-8/T}-3 (219 p+5) e^{-12/T}+\\
&+(375-855 p) e^{-4/T}+18 (25 p-57)\\
D_6 &= -60 (3 p+5) e^{-4/T}+24 (13 p-5) e^{-8/T}+\\
&+(20-52 p) e^{-12/T}-16 (p-21)\\
E_6 &= (1-p) \left(-15 e^{-12/T}+90 e^{-8/T}+75 e^{-4/T}-150\right)\\
F_6 &= (1-p) \left(6 e^{-12/T}-36 e^{-8/T}+30 e^{-4/T}\right)\\
G_6 &=(1-p) \left(e^{-12/T}-6 e^{-8/T}+15 e^{4/T}-10\right)
\end{split}
\end{aligned}
\end{equation}

and the following equation for the critical line
\begin{equation}
\begin{aligned}
p_6^{*}(T)&=\frac{e^{-12/T}-6 e^{-8/T}-45 e^{-4/T}-270}{-223 e^{-12/T}-966 e^{-8/T}-1485 e^{-4/T}+306}\\
&\underset{T \to \infty}= \frac{5}{37}
\end{aligned}
\end{equation}

\section{\texorpdfstring{Landau approach for $q=5$}{Appendix C}}\label{app3}

The rate equation for $q=5$ and the relevant coefficients are given, respectively, by Eqs. \ref{eq:f5} and \ref{eq:f5c}. To follow the Landau approach, we calculate the effective potential $V_5=-\int F_5dm$ and expand it into a Taylor series around $m=0$, keeping the first three non-zero terms:

\begin{equation}
V(m)=A^L_5(T,p)m^2+B^L_5(T,p)m^4+C^L_5(T,p)m^6.    
\end{equation}

where

\begin{widetext}
\begin{equation}
\begin{aligned}   
A^L_5(T,p) &= \frac{\left[10 (35 p-3) e^{-2/T}+5 (65 p-1) e^{-6/T}+(95 p+1) e^{-10/T}-2 (65 p+47)\right]}{2048}\\
B^L_5(T,p) &= \frac{5 m^4 \left[(65 p-1) e^{-10/T}-5 (p-1) e^{-6/T}+(22-150 p) e^{-2/T}+(90 p-26)\right]}{4096}\\
C^L_5(T,p) &= \frac{m^6 \left[10 (23 p-7) e^{-2/T}-5 (27 p+5) e^{-6/T}+(154-202 p)+(5+43) e^{-10/T}\right]}{3072}\\
\end{aligned}
\end{equation}
\end{widetext}

Solving $A^L_5(T,p)=0$ for $p$ (which is equivalent to Eq. \ref{eq:p5})) and putting the solution into $B^L_5(T,p)$ we get

\begin{widetext}
\begin{equation}
    B^L_5(T)=\frac{5 \left(23 e^{-10/T}-e^{-20/T}+3 e^{-16/T}+34 e^{-12/T}+50 e^{-8/T}-57 e^{-6/T}+20 e^{-4/T}-146 e^{-2/T}+74\right)}{128 \left(-19 e^{-10/T}-65 e^{-6/T}-70 e^{-2/T}+26\right)}
\end{equation}
\end{widetext}

Let us first note that $B^L_5=0$ when $T \to \infty$. Secondly, if we denote $\beta = T^{-1}$ in the equation above and then expand Taylor series around $\beta = 0$ and restrict to the first term, we obtain
\begin{equation}
    B^L_5(T)\approx-\frac{5}{32T} < 0
\end{equation}
Taking into account that $A_5^L$ has to be 0 at the critical line and that $C^L_5(T,p_5^{*}(T))>0$ we conclude that the bistability region in the $q=5$ case disappears only when $T \to \infty$.

\section{\texorpdfstring{Analytical formula for the crtitical line $p^*_q(T)$}{Appendix D}}\label{app4}
In this section, we derive a fully analytical formula for the critical line $p^*_q(T)$ given by Eq. (\ref{eq:sep1}) in terms of special functions (Gaussian hypergeometric functions), and then we prove that the limiting case $p^*_q(T \to \infty)$ coincides with the result obtained by Nyczka et al. for the critical value of noise in the monolayer q-voter model \cite{Nyczka2012}.

Let us start by restricting ourselves to only even values of $q$ -- in this way, without losing the generality, we are able to facilitate further calculations. Using this assumption, we are able to notice that the form of $E_{qk}=\min\{1,\exp[2(q-2k)/T]\}$ implies immediately that $E_{ek}=1$ once $k \le q/2$. This observation allows us to separate sums in the nominator and the denominator in Eq. (\ref{eq:sep1}) and rewrite it as

\begin{widetext}
\begin{equation}
p^*_q(T)=\frac{\sum\limits_{k=0}^{q/2}\tbinom{q}{k}\left(2q-2k-1\right)+\sum\limits_{k=q/2+1}^{q}\tbinom{q}{k}\left(2q-2k-1\right)E_{qk}}{\sum\limits_{k=0}^{q/2}\tbinom{q}{k}[(2^{q-1}-1)(1+2k-q)+q]+\sum\limits_{k=q/2+1}^{q}\tbinom{q}{k}[(2^{q-1}-1)(1+2k-q)+q]E_{qk}}.
\label{eq:appsep1}
\end{equation}
\end{widetext}

Let us now observe that the following summations hold true:

\begin{equation}
\begin{aligned}
\begin{split}
\sum\limits_{k=0}^{q/2}\dbinom{q}{k} &= \frac{1}{2} \left[\binom{q}{\frac{q}{2}}+2^q\right]\\
\sum\limits_{k=0}^{q/2}\dbinom{q}{k}k &= 2^{q-2}q\\
\sum\limits_{k=\frac{q}{2}+1}^{q}\dbinom{q}{k}E_{kq} &=e^{-4/T} \binom{q}{\frac{q}{2}+1} \, F1\\
\sum\limits_{k=\frac{q}{2}+1}^{q}\dbinom{q}{k}kE_{kq} &=\frac{1}{2} (q+2) e^{-4/T} \binom{q}{\frac{q}{2}+1} F1+\\&+e^{-8/T} \binom{q}{\frac{q}{2}+2} \, F2
\end{split}
\end{aligned}
\label{eq:appsum}
\end{equation}

where, for brevity, we have used the following notation
\begin{equation}
\begin{aligned}
F1 &={}_2F_1\left(1,1-\frac{q}{2};\frac{q}{2}+2;-e^{-4/T}\right)\\
F2 &={}_2F_1\left(2,2-\frac{q}{2};\frac{q}{2}+3;-e^{-4/T}\right)
\end{aligned}
\end{equation}
to denote relevant Gaussian hypergeometric functions $\, _2F_1(a,b;c;z)$. 
The first two sums in Eq.~(\ref{eq:appsum}) are straightforward and can be calculated by noticing that $\sum_{k=0}^{k=q}\binom{q}{k}=2^q$ and $\sum_{k=0}^{k=q}\binom{q}{k}k=2^{q-1}q$ and then making use of the fact that binomial coefficients are symmetrical. 

The last two sums in Eq.~(\ref{eq:appsum}) can be derived using the series expansion of the Gaussian hypergeometric function $\, _2F_1(a,b;c;z)=\sum _{k=0}^{\infty } \frac{z^k (a)_k (b)_k}{k! (c)_k}$ (see 15.1.1 in \cite{Abramowitz1964}), where $(x)_k$ is the rising factorial, i.e., $(x)_k = \prod_{k=1}^n(x-k+1)$. If $a$ or $b$ is a negative integer $-m$, the infinite series reduces to a polynomial 
\begin{equation}
_2F_1(a,-m;c;z)=\sum _{k=0}^{m} \frac{z^k (-m)_k (b)_k}{k! (c)_k}    
\end{equation} (see 15.4.1 in \cite{Abramowitz1964} and note that $_2F_1(a,b;c;z)={}_2F_1(b,a;c;z)$). Taking into account relations Eq.~(\ref{eq:appsum}) we are able to express $p^*_q(T)$ in a more concise form


\begin{widetext}
\begin{equation}
p^*_q(T)=\frac{4 (q-3) e^{-4/T} \binom{q}{\frac{q}{2}+1} \, F1-8 e^{-8/T} \binom{q}{\frac{q}{2}+2} \, F2+2 \left((2 q-1) \binom{q}{\frac{q}{2}}+2^q (q-1)\right)}{2 \left(2 q+3\ 2^q-6\right) e^{-4/T} \binom{q}{\frac{q}{2}+1} F1+4 \left(2^q-2\right) e^{-8/T} \binom{q}{\frac{q}{2}+2} \, F2+\left(2^q-2-\left(2^q-4\right) q\right) \binom{q}{\frac{q}{2}}+2^q \left(2 q+2^q-2\right)}
\label{eq:apppq}
\end{equation}
\end{widetext}

Let us underline here that special functions used in this derivation are simply another way to express specific sums in Eq. (\ref{eq:appsum}). To obtain the limiting case $p^*_q(T\to\infty)$ let us note that we need to evaluate $\underset{T \to \infty}{F1}$ and $\underset{T \to \infty}{F2}$, i.e., $_2F_1\left(1,1-q/2;q/2+2;-1\right)$ and $_2F_1\left(1,2-q/2;q/2+3;-1\right)$. We shall use the following identities \cite{Abramowitz1964,hyper}

\begin{widetext}
\begin{equation}
\begin{aligned}
    \, _2F_1(a,b;a-b+2;-1)&=\frac{\left(\sqrt{\pi } \Gamma (a-b+2)\right) \left(\frac{1}{\Gamma \left(\frac{a}{2}\right) \Gamma \left(\frac{a+3}{2}-b\right)}-\frac{1}{\Gamma \left(\frac{a+1}{2}\right) \Gamma \left(\frac{a}{2}-b+1\right)}\right)}{2^a (b-1)}\\
    \, _2F_1(a,b;a-b+3;-1)&=\frac{\left(\sqrt{\pi } \Gamma (a-b+3)\right) \left(-\frac{a}{\Gamma \left(\frac{a}{2}+1\right) \Gamma \left(\frac{a+3}{2}-b\right)}+\frac{a (a+1)}{4 \left(\Gamma \left(\frac{a+3}{2}\right) \Gamma \left(\frac{a}{2}-b+2\right)\right)}+\frac{1}{\Gamma \left(\frac{a+1}{2}\right) \Gamma \left(\frac{a}{2}-b+1\right)}\right)}{2^a ((b-1) (b-2))}
\end{aligned}
\end{equation}
\end{widetext}
where $\Gamma(x)$ is the Euler gamma function, which allow us to write that
\begin{equation}
\begin{aligned}
\underset{T \to \infty}{F1} &=\left(\frac{1}{q}+\frac{1}{2}\right) \left(\frac{2^q}{\dbinom{q}{\frac{q}{2}}}-1\right)\\
\underset{T \to \infty}{F2} &=\frac{(q+2) (q+4) \left((q+2) \dbinom{q}{\frac{q}{2}}-2^{q+1}\right)}{4 (q-2) q \dbinom{q}{\frac{q}{2}}}
\end{aligned}
\label{eq:appf}
\end{equation}
After substituting $F1$ and $F2$ in Eq. (\ref{eq:apppq}) with limiting values from Eq. (\ref{eq:appf}) as well as taking the limit of $e^{-4/T}$ and $e^{-8/T}$ and some rudimentary algebra we finally arrive at
\begin{equation}
p^*_q(T \to \infty)=\frac{2 (q-1)}{2(q-1)+2^q}
\end{equation}
that matches the result obtained by Nyczka et al. \cite{Nyczka2012}.

\bibliography{ref}
\end{document}